\documentclass[10pt,aps,twocolumn,floatfix,showpacs,showkeys,preprintnumbers,nofootinbib,superscriptaddress,longbibliography,bibnotes]{revtex4-1}
\pdfoutput=1
\usepackage[colorlinks=true,breaklinks=true]{hyperref}
\usepackage[utf8]{inputenc}
\hypersetup{allcolors=[rgb]{0.0 0.0 0.6},linkcolor=[rgb]{0.75 0.05 0.05}}
\usepackage{amsmath,amssymb}
\usepackage{epsfig}  
\usepackage{graphicx}   
\usepackage{slashed}             
\usepackage{url}
\usepackage{color}
\usepackage{multirow}
\usepackage{placeins}
\usepackage[dvipsnames]{xcolor}
\usepackage{letltxmacro}
\LetLtxMacro{\oldcite}{\cite}
\renewcommand{\cite}[1]{\mbox{\oldcite{#1}}}

\allowdisplaybreaks

\bibpunct{[}{]}{,}{n}{}{,}

\newcommand{\lam}{\Lambda}
\newcommand{\Ham}{{\cal H}}

\begin{document}

\title{A few moments to diagnose fast flavor conversions of supernova neutrinos}

\author{Basudeb Dasgupta}
\email{bdasgupta@theory.tifr.res.in}
\affiliation{Tata Institute of Fundamental Research,
             Homi Bhabha Road, Mumbai, 400005, India.}
\author{Alessandro Mirizzi}
\email{alessandro.mirizzi@ba.infn.it }
\affiliation{Dipartimento Interateneo di Fisica ``Michelangelo Merlin'', Via Amendola 173, 70126 Bari, Italy.}
\affiliation{Istituto Nazionale di Fisica Nucleare - Sezione di Bari,
Via Amendola 173, 70126 Bari, Italy.}
\author{Manibrata Sen}
\email{manibrata.sen@gmail.com }  
\affiliation{Tata Institute of Fundamental Research,
             Homi Bhabha Road, Mumbai, 400005, India.}

\preprint{TIFR/TH/18-08}
\date{July 9, 2018}

\begin{abstract}
Neutrinos emitted from a supernova may undergo flavor conversions almost immediately above the core, with possible consequences for supernova dynamics and nucleosynthesis. However, the precise conditions for such fast conversions can be difficult to compute and require knowledge of the full angular distribution of the flavor-dependent neutrino fluxes, that is not available in typical supernova simulations. In this paper, we show that the overall flavor evolution is qualitatively similar to the growth of a so-called ``zero mode'', determined by the background matter and neutrino densities, which can be reliably predicted using only the second angular moments of the electron lepton number distribution, i.e., the difference in the angular distributions of $\nu_e$ and $\bar{\nu}_e$ fluxes. We propose that this zero mode, which neither requires computing the full Green's function nor a detailed knowledge of the angular distributions, may be useful for a preliminary diagnosis of possible fast flavor conversions in supernova simulations with modestly resolved angular distributions.
\end{abstract}

\maketitle

\section{Introduction}
\label{sec:1}
The interior of a supernova (SN) hosts a unique laboratory to probe quantum correlations between neutrinos. For instance, at distances $r \lesssim {\mathcal O}(10^2)$\,km from the centre of the SN, the neutrino density $n_{\nu}$ is so high that they themselves produce a collective potential $\mu=\sqrt{2}G_F\, n_{\nu_e}$, defined in terms of the electron neutrino density $ n_{\nu_e}$. This potential, being much larger than the neutrino oscillation frequency in vacuum, $\omega_{\rm vac}=\Delta m^2/(2E)$ for a typical neutrino energy $E$, leads to correlated neutrino flavor evolution~\cite{Pantaleone:1992eq,Kostelecky:1994dt,Pastor:2002we}. The past decade of research in this area has unearthed many fascinating features in the collective oscillations of neutrinos~\cite{Sawyer:2005jk, Duan:2006an, Hannestad:2006nj, Fogli:2007bk, Sawyer:2008zs, Dasgupta:2009mg, Sawyer:2015dsa}. See refs.\,\cite{Duan:2010bg,Mirizzi:2015eza,Chakraborty:2016yeg,Horiuchi:2017sku} for recent reviews.

In a series of papers~\cite{Sawyer:2005jk,Sawyer:2008zs,Sawyer:2015dsa}, Ray Sawyer has pointed out a new mechanism for self-induced flavor conversions called ``fast'' instabilities. These are expected to develop at very short distances,  $r \lesssim {\mathcal O}(1)$\,m, from the neutrinosphere and grow with a rate $\mu$, i.e., not only faster than the usual neutrino oscillations but also than the relatively slower collective oscillations, that lead to spectral splits and swaps~\cite{Fogli:2007bk,Raffelt:2007cb,Dasgupta:2008cd,Dasgupta:2009mg}, growing at a rate $\sqrt{\omega_{\rm vac}\,\mu}$~\cite{Hannestad:2006nj}. Recently, several groups have confirmed these results and further developed the original insights~\cite{Chakraborty:2016lct,Dasgupta:2016dbv,Izaguirre:2016gsx,Capozzi:2017gqd, Dighe:2017sur, Dasgupta:2017oko,Abbar:2017pkh}. In particular, it has been understood that a necessary condition for fast conversions is that there is a ``crossing" in the electron lepton number (ELN) angular distribution, i.e., the difference of the $\nu_e$ and $\bar{\nu}_e$ densities must change its sign as a function of emission angle. This crossing condition is  similar to how collective spectral swaps require a crossing in the energy spectrum~\cite{Dasgupta:2009mg}. In the neutrino decoupling region inside a SN, where the different flavors have significantly different angular distributions, a crossing in the angular spectrum could be present. As a result, fast conversions may occur and lead to potentially radical changes in SN dynamics and neutrino signals. 

The possibility of fast conversions need to be explored systematically in SN simulations. First steps in this direction were taken recently~\cite{Tamborra:2017ubu}, where a dedicated analysis of the angular distributions of the neutrino radiation field  for several spherically symmetric (1D)
supernova simulations has not found any crossing  in the ELN near the neutrinosphere. Ref.\,\cite{Abbar:2017pkh}, on the other hand, found an instability in a $8.8\,M_\odot$ electron capture SN simulation by the Garching group. More generally, 2D or 3D models can exhibit Lepton-Emission Self-sustained Asymmetry (LESA)~\cite{Tamborra:2014aua}, i.e., a large-scale dipole in the ELN emission, which also makes a crossing more likely to occur. Unfortunately, a study of fast oscillations in 2D  or 3D simulations has been lacking for two reasons. Firstly, the study of fast neutrino instabilities requires characterizing the singularities of the Green's
function of the system~\cite{Capozzi:2017gqd}. This is a computationally demanding task even for the simplest toy models, and perhaps prohibitively difficult for the multidimensional continuous angular distributions found in SN simulations. Secondly, most of the state-of-the-art simulations~\cite{Tamborra:2014aua,Bruenn:2014qea, OConnor:2015rwy, Nagakura:2017mnp, Richers:2017awc, Vartanyan:2018xcd, Just:2018djz, Cabezon:2018lpr, Pan:2018vkx}, maintain only the moments of the angular distributions of fluxes, and not the full distributions. This lack of information seems to preclude even a linear stability analysis that requires knowing these distributions. One may in fact worry, whether these coarse-grained distributions can correctly capture the physics of fast oscillations. Therefore, it is necessary to consider an alternative approach that uses available simulations in an optimal fashion.

In this work, we propose a simple analytical tool to diagnose fast instabilities. Our proposal is based on identifying a specific Fourier mode of the flavor instability field, that we call the ``zero mode''. The growth rate of this mode, calculated from the stability analysis, crudely approximates the growth of flavor conversions in detailed numerical calculations, essentially all the way until the instability saturates. The zero mode has an easily calculable growth rate, that depends only on the second moments of the ELN. Thus, the proposed method has the dual advantage of not requiring complete knowledge of the neutrino distributions and being computationally far less expensive compared to a full-fledged numerical solution or a full characterization of the Green's function. In fact, in the absence of more detailed knowledge of the ELN distributions, as appears to be the case for available 2D and 3D SN simulations, this may be the only practical recourse to look for possible instabilities. Therefore, we expect that this method will be useful to scan the different regions of a SN in multidimensional simulations and study the possibility of fast flavor conversions therein. 

We discuss these issues in the following sections. In Sec.\,\ref{sec:2}, we write down the equations of motion (EoMs) and review the framework for studying fast neutrino oscillations. In Sec.\,\ref{sec:3} we present our method for diagnosing instabilities and in Sec.\,\ref{sec:4} perform numerical tests of the same, for simple box-like angular distributions for the $\nu_e$ and $\bar{\nu}_e$ as well as realistic angular distributions inspired by 1D SN models. Finally, in Sec.\,\ref{sec:5}, we conclude with a brief summary.

\section{Equations of Motion}
\label{sec:2}
Neglecting collisions, the dynamics of $\varrho_{\bf p}$, the matrices of neutrino phase space occupation number densities  for the momentum ${\bf p}$,  is described by the following EoMs\,\cite{Sigl:1992fn, Strack:2005ux, Vlasenko:2013fja, Volpe:2015rla,Hansen:2016klk, Stirner:2018ojk}
\begin{equation}
\partial_t \varrho_{{\bf p}} + {\bf v}_{\bf p} \cdot \nabla \varrho_{{\bf p}} 
= - i [\Ham_{{\bf p}}, \varrho_{{\bf p}}] 
\,\ ,
\label{eq:eom}
\end{equation}
where, in the Liouville operator on the left-hand side,  the first term accounts for explicit dependence on time $t$, while the second term, proportional to the neutrino velocity ${\bf v_p}$, encodes the dependence on position ${\bf x}$ due to particle free streaming. 
The right-hand-side contains the oscillation Hamiltonian
\begin{equation}
\Ham_{{\bf p}}= \Ham_{{\rm vac}} + \Ham_{\rm mat} + \Ham_{\nu\nu}\, ,
\end{equation} 
where in a two-flavor approximation, one has
\begin{equation}
\Ham_{{\rm vac}}={\rm diag(-\omega_{\rm vac}/2,+\omega_{\rm vac}\,/2)}\,,
\end{equation}
in the mass basis, and
\begin{equation} 
\Ham_{\rm mat}=\sqrt{2}G_F n_e\,{\rm diag(1,0)}\,,
\end{equation}
in the weak interaction basis, contains the refractive effect of charged leptons in the medium, while
\begin{equation} 
\Ham_{\nu\nu} = \sqrt{2} G_F \int {d^3 {\bf q}}/{(2 \pi)^3} ({\varrho_{\bf q}} - {\bar\varrho_{\bf q}}) (1 -{\bf v}_{\bf p}\cdot {\bf v}_{\bf q})
\label{eq:Hnunu} 
\end{equation}
contains the similar effect due to background neutrinos. Antineutrinos are described similarly using $\bar\varrho_{\bf p}$, with $\Ham_{{\rm vac}}$ replaced by $-\Ham_{{\rm vac}}$.


The matrix $\varrho$ can be written in the weak-interaction basis as~\cite{Banerjee:2011fj, Izaguirre:2016gsx, Capozzi:2017gqd, Morinaga:2018aug}
\begin{equation}
\varrho = \frac{f_{\nu_e} + f_{\nu_x}}{2}
\left(\begin{array}{cc} 1 & 0 \\
0 & 1 \end{array} \right)
+  \frac{f_{\nu_e} - f_{\nu_x}}{2} 
\left(\begin{array}{cc} s & S \\
S^* & -s \end{array} \right) \,\ ,
\label{eq:density}
\end{equation}
where ${f_{\nu_e}}$ and ${f_{\nu_x}}$ are the occupation number densities at momentum ${\bf p}$, and $\nu_x$ is the relevant linear combination of $\nu_\mu$ and $\nu_\tau$. Here and onwards, we drop the subscript $\bf p$, which indicated that the $\varrho$ were indexed by their momenta, to lighten the notation.

One focuses on length and time scales over which $f_{\nu_{e}}$ and $f_{\nu_{x}}$ can be taken to be homogeneous and static, and thus the spatial and temporal dependence of $\varrho$ is contained in $S$ and $s$. The complex scalar field $S (t, {\bf x})$ encodes the $\nu_e\nu_x$ mean-field flavor coherence, and measures the extent of flavor conversion. To begin with, the neutrinos are initially in their unoscillated states, hence the initial condition is $S(0,{\bf x})=0$. The real field $s (t, {\bf x})$ encodes flavor occupation number, and satisfies $s^2  +|S|^2=1$, for each momentum $\bf p$.

In the context of fast conversions, the effect of background neutrinos via $\Ham_{\nu\nu}$ far exceeds that of the vacuum Hamiltonian $\Ham_{{\rm vac}}$, which mainly plays the role of generating an initial disturbance to seed the oscillations. Hence, $\Ham_{{\rm vac}}$ can be neglected and the explicit dependence on energy $E$, via $\omega_{\rm vac}$, disappears from the EoMs. The neutrino and antineutrino modes then enter the Hamiltonian in Eq.\,(\ref{eq:Hnunu}) only via the difference of occupation number densities integrated over energy, i.e., the ELN angular distribution~\cite{Chakraborty:2016lct},
\begin{equation}
G_{\bf v} = \sqrt{2} G_F \int_{0}^{\infty}\frac{dE\,E^2}{2 \pi^2}\left[f_{\nu_e}(E,{\bf v})-f_{\bar\nu_e}(E,{\bf v}) \right] \,\ ,
\label{eq:eln}
\end{equation}
where we assume $\nu_x$ and $\bar{\nu}_x$ have identical distributions.

We will often use the ``4-vector'' notation, e.g., $a^{\mu}=(a^0,{\bf a})$, advocated in Ref.\,\cite{Izaguirre:2016gsx}. For the familiar quantities, i.e., position $x^\mu=(t,{\bf x})$, momentum $p^\mu=(E,{\bf p})$, and wavevectors $k^\mu=(\omega,{\bf k})$ and $K^\mu=(\Omega,{\bf K})$, the zeroth component is denoted by its more recognizable symbol instead. The neutrinos are taken to be ultra-relativistic, with $E=|{\bf p}|$, so $v^\mu=(1,{\bf p}/E)$, i.e., the zeroth component of their velocity is 1 and the spatial components are given by a unit vector ${\bf v}={\bf p}/E$. In this notation, one can define a matter current $\Lambda^\mu=\sqrt{2}G_F\,v^\mu_e\,n_e$ and an ELN current $\Phi^\mu=\int d\Gamma\,v^\mu\,G_{\bf v}$, where $d\Gamma = d{\bf v}/(4\pi)$, which contain the effect of $\Ham_{\rm mat}$ and $\Ham_{\nu\nu}$, respectively.

The key feature which determines if the initial flavor composition is unstable and can undergo fast conversions, is if the ELN distribution $G_{\bf v}$ crosses zero as a function of any of its arguments. This essentially requires that the flux of neutrinos is larger than that of antineutrinos in some direction, while being smaller in another direction. If the lepton asymmetry,  $\varepsilon=(n_{\nu_e}- n_{\bar\nu_e})/n_{\nu_e}= \Phi^0/\mu\,$ is small, then the ELN distribution $G_{\bf v}$ can have a crossing. This is because the density of forward-going $\bar\nu_e$ can exceed that of $\nu_e$, between the $\bar{\nu}_e$ and $\nu_e$ neutrinospheres. 
\pagebreak


The onset of the conversions can be examined by linearizing Eq.\,(\ref{eq:eom}), using that initially  $|S| \ll 1$ and \mbox{$s \simeq1$}. Starting from the linearized  equations of motion~\cite{Banerjee:2011fj}, one seeks plane wave solutions obeying~\cite{Dasgupta:2015iia}
\begin{equation}
S_{\bf v}(t,{\bf x}) = Q_{\bf v} e^{-i(\Omega t - {\bf K} \cdot {\bf x})} \,\  .
\label{eq:wave}
\end{equation}
A specific eigenmode of flavor conversion can be labeled by its frequency and wavevector, $\Omega$ and ${\bf K}$, respectively. If there are modes that have a complex $\Omega$, such modes may lead to exponentially growing instabilities.

The currents $\Phi^\mu$ and $\Lambda^\mu$ lead to a common rotation for all modes, which does not lead to any instabilities as such. So it is more convenient to work in a rotating coordinate system where this common rotation is not present. In such a co-rotating frame, the different modes of flavor conversions are labeled by the shifted frequency and wavevectors, 
\begin{eqnarray}
\label{omega_redef}\omega&=&\Omega-(\Lambda^0+\Phi^0)\,,\,{\rm and}\\
\label{k_redef}{\bf k}&=&{\bf K}-({\bf \Lambda}+{\bf \Phi})\,,
\end{eqnarray}
respectively. Note that the frequency and wavevector of the modes in the co-rotating frame, $\omega$ and ${\bf k}$, have the same imaginary parts as in the non-rotating frame. Hence this shift, while simplifying the EoMs, does not give rise to any extra spurious instabilities. Of course, one must be careful of the shift when identifying a specific mode of the flavor conversion field $S$,  e.g., the homogeneous mode, previously labeled by ${\bf K}=0$, now corresponds to the mode ${\bf k}=-({\bf \Lambda}+{\bf \Phi})$.

The  $\omega$ and ${\bf k}$ are related  by the dispersion relation of the system~\cite{Izaguirre:2016gsx}
\begin{equation}
D(\omega,{\bf k})=\textrm{det}\left[\Pi^{\mu \nu} (\omega, {\bf k}) \right] =0 \,\ ,
\label{eq:dispmain}
\end{equation}
where
\begin{equation}
\Pi^{\mu \nu} = \eta^{\mu \nu} +\int d\Gamma\,G_{\bf v} \frac{v^{\mu}v^{\nu}}{\omega-{\bf k} \cdot {\bf v}} \,\ ,
\label{eq:polar}
\end{equation}
with $\eta^{\mu\nu}=\textrm{diag}(+1,-1,-1,-1)$ being the metric tensor. 
In the remainder of the paper we will refer to the ${\bf k}=0$ mode as the ``zero mode''. This mode will be the focus of our work, and we will come back to it in the next section.

We end this section with a brief remark about the role of ordinary matter density. From the definition $\omega=\Omega-(\Lambda^0+\Phi^0)$, one sees that ${\rm Im}(\omega)$ has no dependence on the ordinary matter density encoded in $\lam^0$, which merely leads to a shift in ${\rm Re}(\omega)$, as noted in Refs.~\cite{Dasgupta:2015iia, Capozzi:2016oyk}. Presence of a finite matter density only delays the onset by suppressing the mixing angle, keeping the growth rate same. A nonzero ordinary matter current ${\bf \Lambda}$ on the other hand can change ${\rm Im}(\omega)$, but is expected to be negligible in a SN-like environment where the ordinary matter has small velocity anisotropy. In the remainder of this paper, we will ignore effects of ordinary matter.

\section{Zero Mode and Moments}
\label{sec:3}

Our proposal in this paper is to focus on the zero mode, labeled by ${\bf k}=0$. This is motivated by the fact that the calculation of $\omega$ for this mode is significantly simpler than a full characterization of the roots of the dispersion relations, $D(\omega, {\bf k})$~\cite{Capozzi:2017gqd}. In fact, for this mode the $\omega$ in Eq.\,(\ref{eq:polar}) can be pulled out of the integrals, and Eq.\,(\ref{eq:dispmain}) becomes 
\begin{equation}
D(\omega,0)=\textrm{det}\left(\eta^{\mu\nu}+\frac{1}{\omega}\,V^{\mu\nu}\right)=0\,,
\label{eq:disp0}
\end{equation}
i.e., $D(\omega,0)$ is a polynomial in $\omega$. The specific model of SN neutrino fluxes and their angular distributions, encoded in the ELN, only enters the equation through the tensor  $V^{\mu\nu}$ that contains the second moments of velocity, namely,
\begin{equation}
V^{\mu\nu} = \int d\Gamma\, {v^\mu v^\nu} G_{\bf v}\,\ . \\
\label{eq:momenta}
\end{equation}
This, in turn, depends on the second moments of neutrinos velocities evaluated using the flavor-dependent phase space distributions,  i.e.,
\begin{equation}
V^{\mu\nu}= \langle {v^\mu v^\nu} \rangle_{\nu_e} - \langle {v^\mu v^\nu} \rangle_{\bar{\nu}_e}\,, 
\end{equation}
where the notation $\langle \ldots \rangle_{\nu_\alpha}$ refers to
\begin{equation}
\langle \ldots \rangle_{\nu_\alpha} \equiv \sqrt{2} G_F \int \frac{d^3{\bf p}}{(2\pi)^3}\, (\ldots)\,f_{\nu_\alpha}({\bf p})\,.
\end{equation}

\begin{figure*}[!t]
\begin{centering}
\hspace{0.2in}\includegraphics[width=0.8\columnwidth,height=0.66\columnwidth]{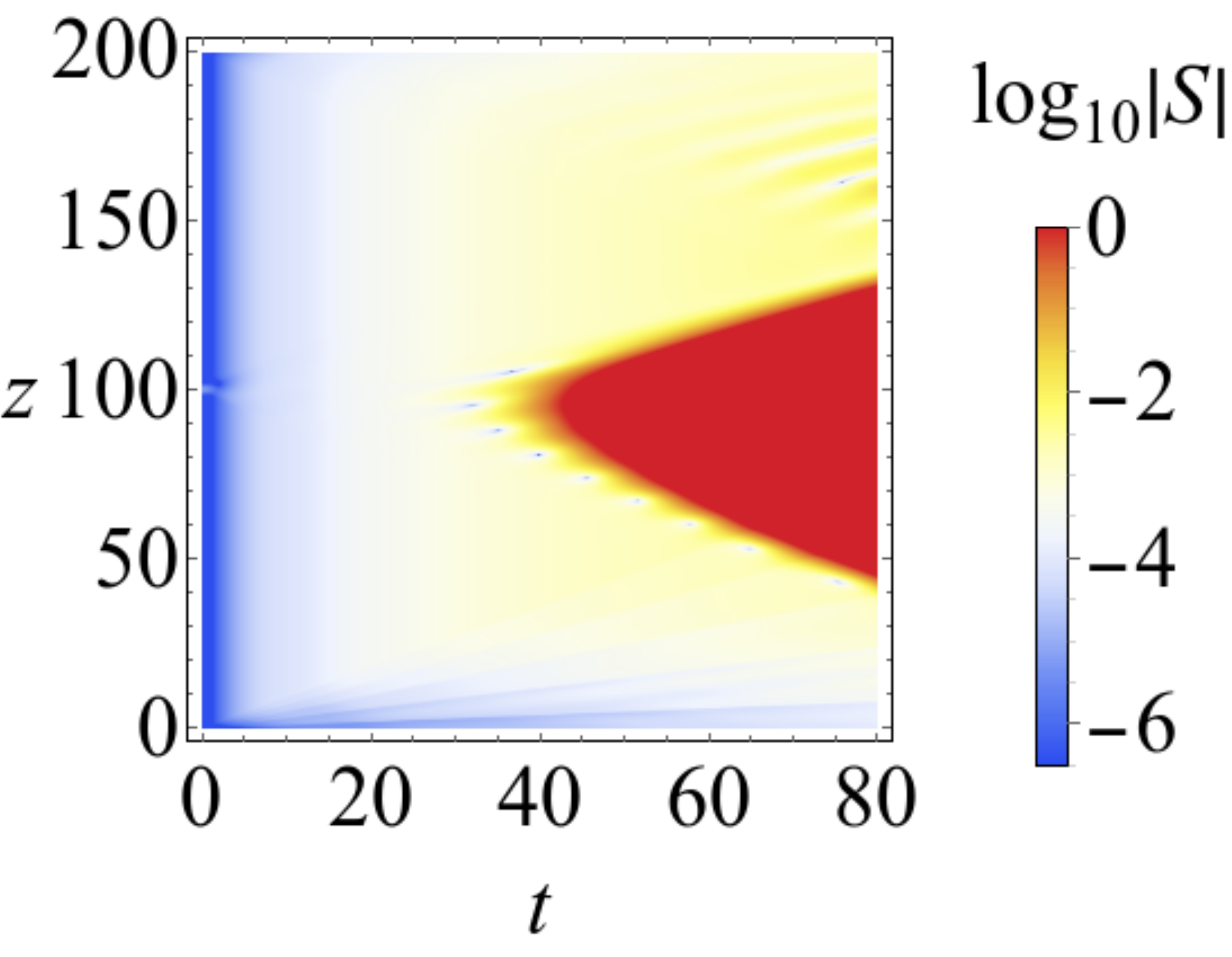}\,
\hspace{1.0cm}
\includegraphics[width=0.95\columnwidth]{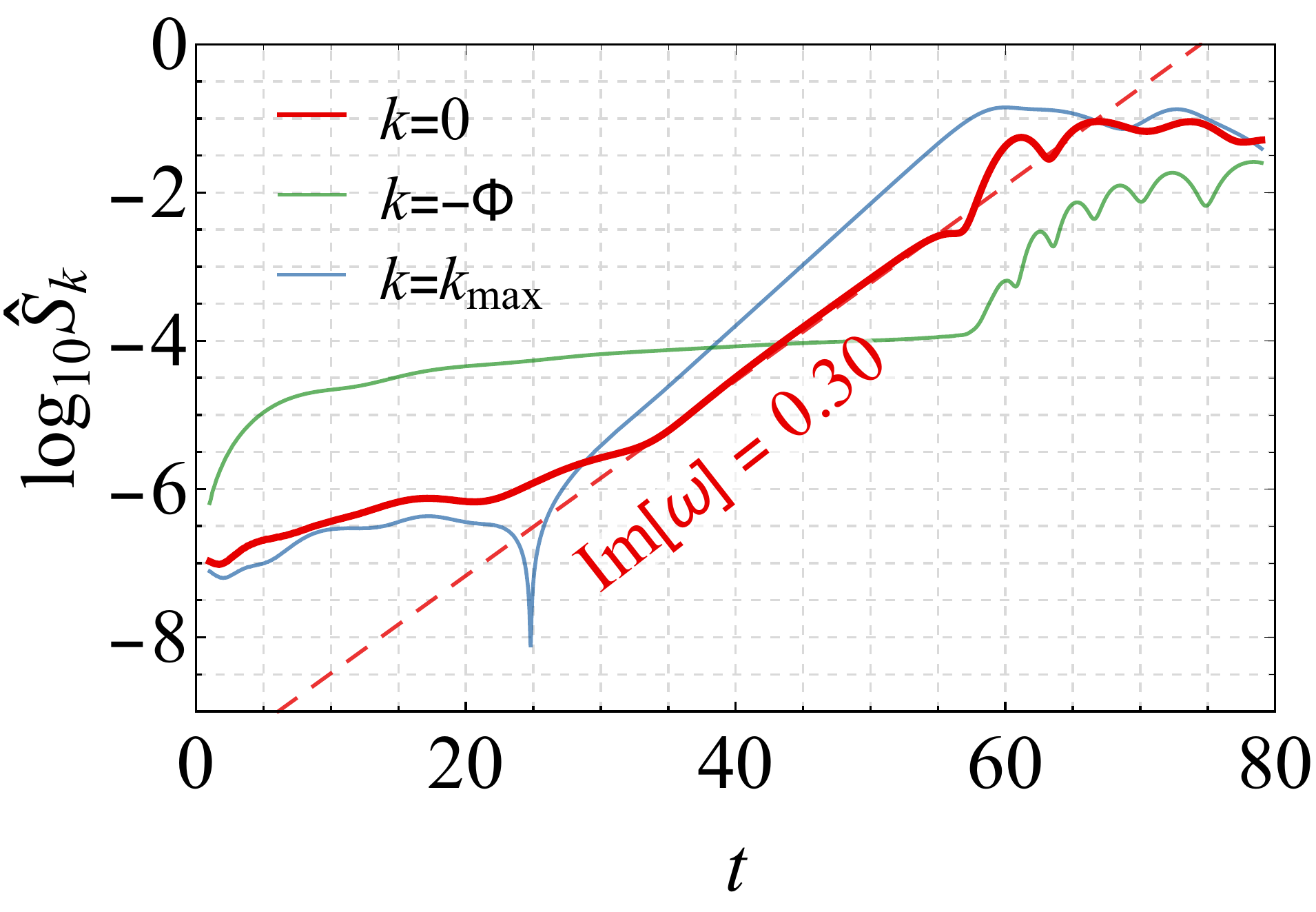}
\end{centering}
\caption{Growth of absolute flavor instability for a toy model of neutrinos with a box-like ELN distribution with a crossing at $v_{\rm c}=0$, given by $(G_{\nu_e}, G_{\bar\nu_e})=(0.3,0.5)$ for $v<v_{\rm c}$ and \mbox{$(G_{\nu_e}, G_{\bar\nu_e})=(1.2,0.5)$} for $v>v_{\rm c}$. \emph{Left panel:} The instability is absolute, and spreads around its origin without drifting. \emph{Right panel:} The numerically computed growth rate of the zero mode labeled by $k = 0$ (continuous red line), the true homogeneous mode labeled by $k = -\Phi$ (unbroken green line), and the mode with the largest growth rate labeled by $k = k_{\rm max}$ (unbroken blue line). The numerically observed growth rate for the zero mode matches the analytical prediction using the moments (dashed red line).} 
\vspace{0.4cm}
\label{fig:boxabsol}
\end{figure*}  

For spherically symmetric SN simulations, which are effectively 1D, Eq.\,(\ref{eq:disp0}) is explicitly quadratic in $\omega\,,$ 
\begin{equation}
\bigl(\omega+V^{00}\bigr)\bigl(\omega-V^{11}\bigr)+\bigl(V^{01}\bigr)^2=0\,\,,
\label{eq:qeq}
\end{equation}
with the solution
\begin{equation}
\omega=\frac{1}{2}\biggl(V^{11}-V^{00}\pm\sqrt{\left(V^{00}+V^{11}\right)^2-4\left(V^{01}\right)^2}\biggr)\,.
\end{equation}
The condition for the zero mode to become unstable is simply that the discriminant become negative, i.e.,
\begin{equation}
\Delta=\left(V^{00}+V^{11}\right)^2-4\left(V^{01}\right)^2 <0\, .
\label{eq:discond}
\end{equation}
If this condition is satisfied, the mode grows at a rate
\begin{equation}
{\rm Im}(\omega)=\frac{1}{2}\left[4\left(V^{01}\right)^2 - \left(V^{00}+V^{11}\right)^2\right]^{1/2}\,.
\label{eq:growthprox}
\end{equation}
For multidimensional models, i.e., for 2D or 3D simulations, Eq.\,(\ref{eq:disp0}) is a cubic or quartic equation for $\omega$, respectively. In either case, the instability growth rate is similarly calculable as the imaginary part of $\omega$, if the flavor-dependent second moments $\langle {v^\mu v^\nu} \rangle_{\nu_\alpha}$ are known.
\pagebreak

We propose that one should check for fast instabilities in a SN simulation, by testing the condition in Eq.\,(\ref{eq:discond}) locally in each simulation cell. It is our understanding that SN simulations often do not track the complete distribution $f(E,{\bf v})$. This precludes an exhaustive search for fast instabilities by studying the solution of the dispersion relation. However, one can learn about the stability of the zero mode without such detailed information. If the ELN distribution $G_{\bf v}$ is spherically symmetric, the tensor $V^{\mu\nu}$ has no cross terms and only depends on $\langle 1 \rangle_{\nu_\alpha}$, $\langle v_r \rangle_{\nu_\alpha}$, and $\langle v_r^2 \rangle_{\nu_\alpha}$, i.e., the zeroth, first, and second moments of the radial velocity.  Such information is readily available even in the spherically symmetric SN simulations and a search for instabilities using Eq.\,(\ref{eq:discond}) is straightforward. In general, the  terms like $\langle {v^\mu v^\nu} \rangle_{\nu_\alpha}$ are important. Some multidimensional SN simulations can provide these cross moments and may allow one to search for fast instabilities. In these cases, if Eq.\,(\ref{eq:disp0}) has complex solutions for $\omega$ in some region in a SN simulation, it indicates that fast conversions should occur.

Finally, we note that, besides the zero mode we have identified, there are two other important modes. The true homogeneous mode of the system is given by ${\bf K}=0$, which is now labeled by ${\bf k}=-({\bf \Lambda}+{\bf \Phi})$. This mode, that has conventionally been studied in calculations that enforce an evolution in time and ignore spatial variations, need not have an instability as will be clear from some of the examples we study in the next section. Nonetheless, our method cannot be used to predict the behavior of this mode. On the other hand, the mode with the maximum growth rate can be determined using the condition that the group velocity of that flavor wave is zero, i.e., $\partial \omega/\partial k|_{k_{\rm max}}=0$~\cite{Capozzi:2017gqd}. Predicting this mode and its growth rate also requires knowing the full dispersion relation. Although our method is not useful to study these modes directly, we will find that the exponential growth of the zero mode, accurately predicted by Eq.\,(\ref{eq:growthprox}), is a good proxy for the overall flavor evolution.

\begin{figure*}[!t]
\begin{centering}
\hspace{0.1in}\includegraphics[width=0.8\columnwidth,height=0.7\columnwidth]{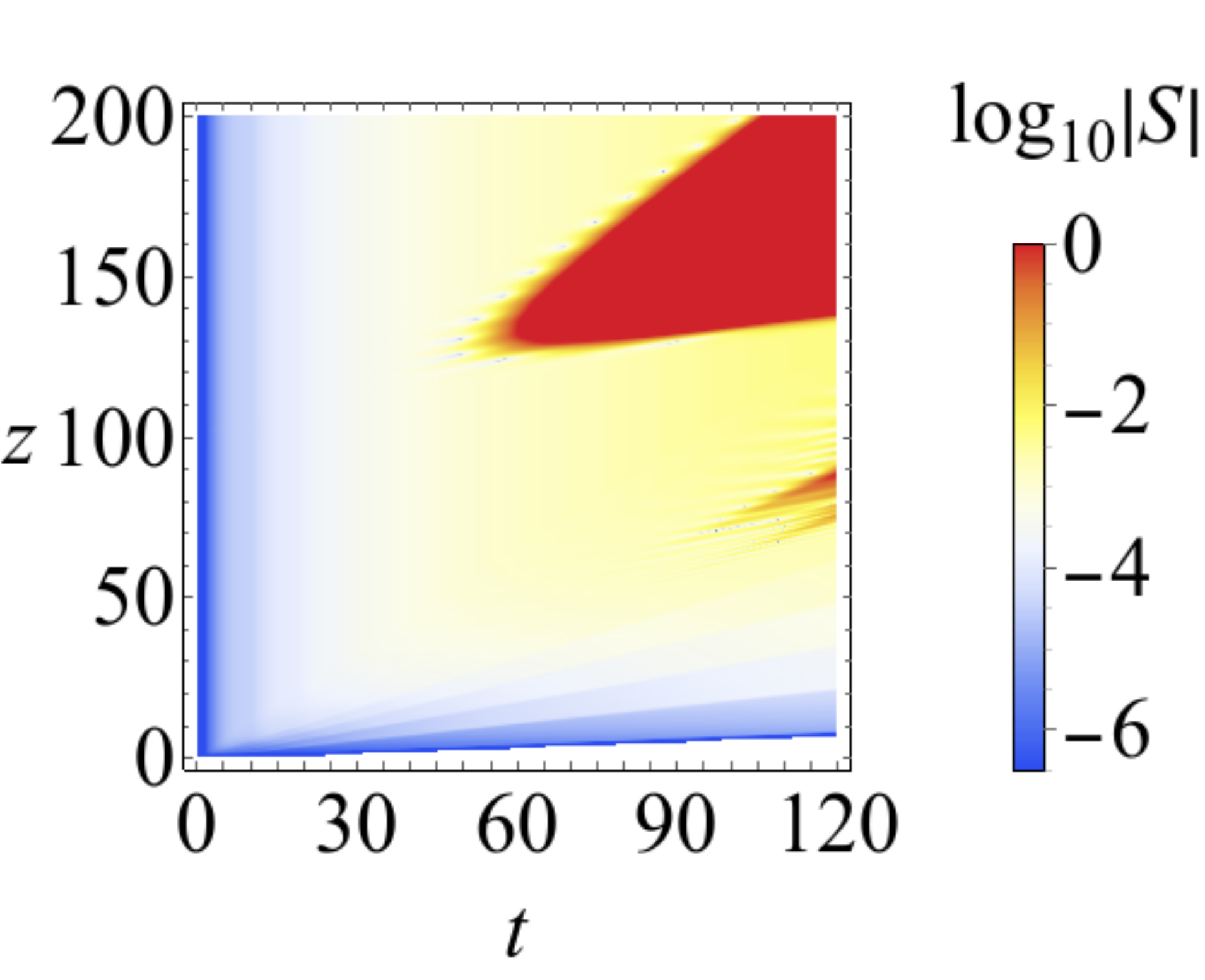}\,
\hspace{1.0cm}
\includegraphics[width=0.95\columnwidth]{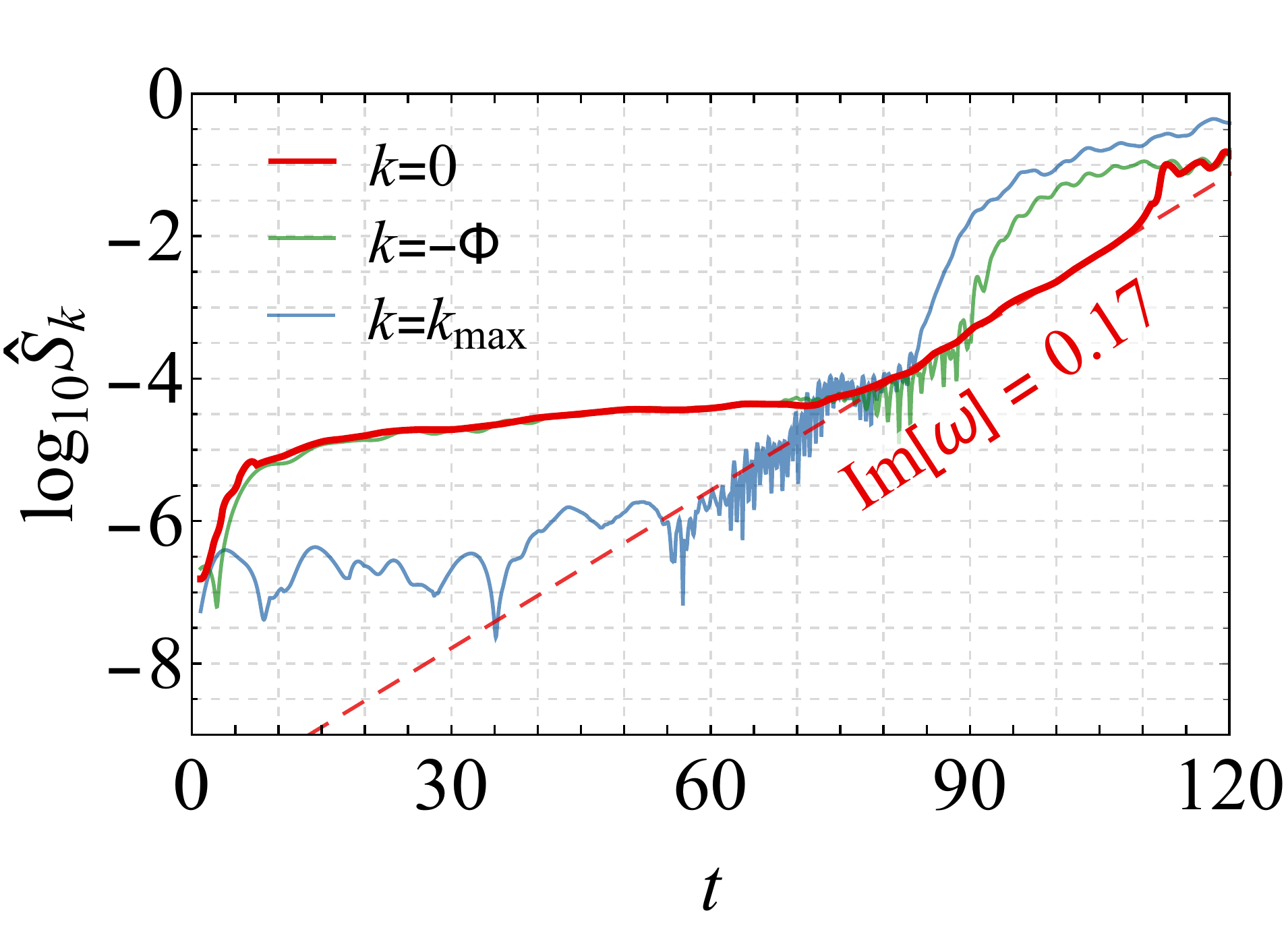}
\end{centering}
\caption{Growth of convective flavor instability for a toy model of neutrinos with a box-like ELN distribution, given by $(G_{\nu_e}, G_{\bar\nu_e})=(0.6,1.0)$ for $0<v< v_{\rm c}$ and $(G_{\nu_e}, G_{\bar\nu_e})=(2.4,1.0)$ for $v_{\rm c}<v<1.0$, with $v_{\rm c}=0.5$. \emph{Left panel:} The instability grows along a particular direction, spreading around it and advecting away from the original site of instability at $z=100$. \emph{Right panel:} The numerically computed growth rate of the zero mode labeled by $k = 0$ (continuous red line), the true homogeneous mode labeled by $k = -\Phi$ (unbroken green line), and the mode with the largest growth rate labeled by $k = k_{\rm max}$ (unbroken blue line). The numerically observed growth rate for the zero mode agrees with the analytical prediction using the moments (dashed red line).} 
\label{fig:boxconv}
\end{figure*}  

\section{Numerical tests in 1D}
\label{sec:4}

In this section, we demonstrate our proposed method using the flavor evolution of models with continuous ELN, in one spatial dimension $z$ and time $t$, neglecting ordinary matter (i.e., $\Lambda=0$). In one spatial dimension, the ${\bf k}$ and ${\bf \Phi}$ vectors can be labeled by their magnitudes $k$ and $\Phi$, respectively. We numerically solve the nonlinear partial differential EoMs, i.e., Eq.\,(\ref{eq:eom}), for several models and compare the flavor evolution, thus obtained, with the growth rate predicted by the corresponding moments. First, we will consider a few toy examples with box-like angular distributions for the $\nu_e$ and $\bar\nu_e$, with a crossing at $v=v_{\rm c}$, and then show a calculation with more realistic distributions inspired by SN simulations. 

 We work in units such that the neutrino potential $\mu=1$, and times and lengths are expressed in units of $\mu^{-1}$. In vacuum, the  oscillation frequency is taken to be $\omega_{{\rm vac}} =9 \times 10^{-5}$ while the effective mixing angle is $\vartheta=10^{-3}$. We assume an inverted neutrino mass ordering, but the results are insensitive to this choice. The solution is found over the $z$-$t$ plane, allowing $z$ to take values in the interval $(0:z_{\rm max})$ and $t$ in $(0:t_{\rm max})$. The boundary conditions are chosen such that flavor-pure modes are emitted at all $z$ when $t=0$, with $S(z,0)$ being a Gaussian wavepacket centered around $z=100$ with small width $\sigma=1$ and initial amplitude $10^{-6}$ for both the real and imaginary part of $S$. Initially, the $k=0$ mode peaks, and the seeds for all other  $k$ modes are smaller.

As the first case, we take a box-like distribution given by \mbox{$(G_{\nu_e}, G_{\bar\nu_e})=(0.3,0.5)$} for $-1<v<0$ and $(G_{\nu_e}, G_{\bar\nu_e})=(1.2,0.5)$ for $0<v<1$. The ELN, $G_{\nu_e}- G_{\bar\nu_e}$, changes sign and presents a crossing at $v_{\rm c}=0$. In this case, there are counter-going neutrinos, and one expects the instability to be absolute, spreading around its origin without drifting~\cite{Capozzi:2017gqd}.

In the left panel of Fig.\,\ref{fig:boxabsol} we show the numerical evolution of $|S(z,t)|$, in the $z$-$t$ plane. This is obtained by numerically solving the nonlinear partial differential EoMs for the model. An instability corresponds to a growth of $|S(z,t)|$. Here, we see that an absolute instability is generated at $t \simeq 30$ and gradually spreads over space without drifting. The nonlinear regime is reached at $t \simeq 60$, when $|S(z,t)| \sim {\mathcal O} (10^{-1})$. 

In order to  estimate the temporal growth of the instability,  we look at the Fourier transform of $S(z,t)$ as a function of $t$,
\begin{equation}
{\hat S}_K(t) = \frac{1}{z_{\rm max}}\int_{0}^{z_{\rm max}} dz\, e^{i\,K z} \,S(z,t)\,\ ,
\end{equation}
The $S(z,t)$ we obtain from the numerical solution of the EoMs is not in the co-rotating frame, and its Fourier modes correspond to the unshifted wavenumbers labeled by $K$, as denoted above. However, we will continue to work in the co-rotating frame and relabel the modes using $k=K-\Phi$ to obtain
\begin{equation}
{\hat S}_k(t) = \frac{1}{z_{\rm max}}\int_{0}^{z_{\rm max}} dz\, e^{i\,\left(k+\Phi\right)z} \,S(z,t)\,\ .
\label{Fourier}
\end{equation}
We remind, the zero mode is labeled by $k=0$, the homogeneous mode by $k=-\Phi$, and the mode with maximum growth by $k=k_{\rm max}$.

In the right panel of Fig.\,\ref{fig:boxabsol} we plot ${\log}_{10} {\hat S}_{k}(t)$ versus $t$ as obtained from the numerical data for different $k$ modes. The zero mode, labeled by $k=0$ (continuous red line) shows an initial slow growth until $t \lesssim 30$, followed by an exponential rise until $t \gtrsim 60$, after which the conversions saturate. The initial slow phase has been identified as an ``onset'' phase and depends logarithmically on the initial seed~\cite{Dasgupta:2017oko}. 
 In the regime with exponential growth, the agreement between the numerical solution  and the analytical prediction of the growth rate (dashed red line) is excellent, both showing a growth rate ${\rm Im}(\omega) = 0.30$. For comparison, we also show the true homogeneous mode, given by $k=-\Phi$ (unbroken green line) and the mode with the largest growth, given by $k=k_{\rm max}$ (unbroken blue line). Note that in this example, the homogeneous mode does not have a linear instability. It is marked by a larger seed and a longer onset period, before non-linearity sets in. The $k=k_{\rm max}$ mode, on the other hand, clearly has a larger growth rate than the zero mode and becomes non-linear earlier.  However, there is no simple analytical expression for the growth rate of this mode. One needs to find the complete dispersion relation, setting $k=k_{\rm max}$ in Eq.\,(\ref{eq:dispmain}). We have identified this mode by numerically searching for the highest growth rate across all $k$.

\begin{figure*}[!t]
\begin{centering}
\includegraphics[width=0.8\columnwidth]{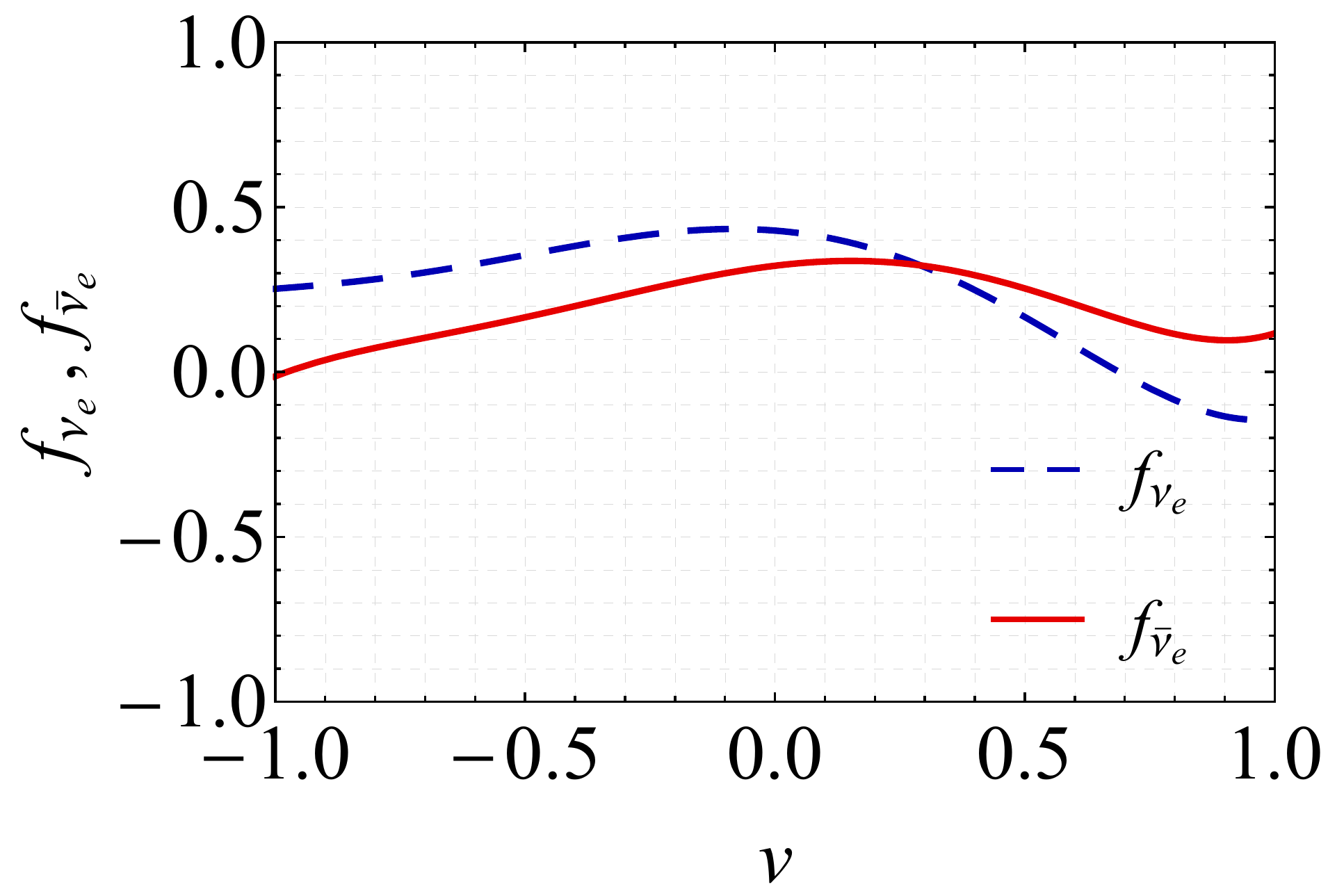}
\hspace{1.4cm}
\includegraphics[width=0.8\columnwidth]{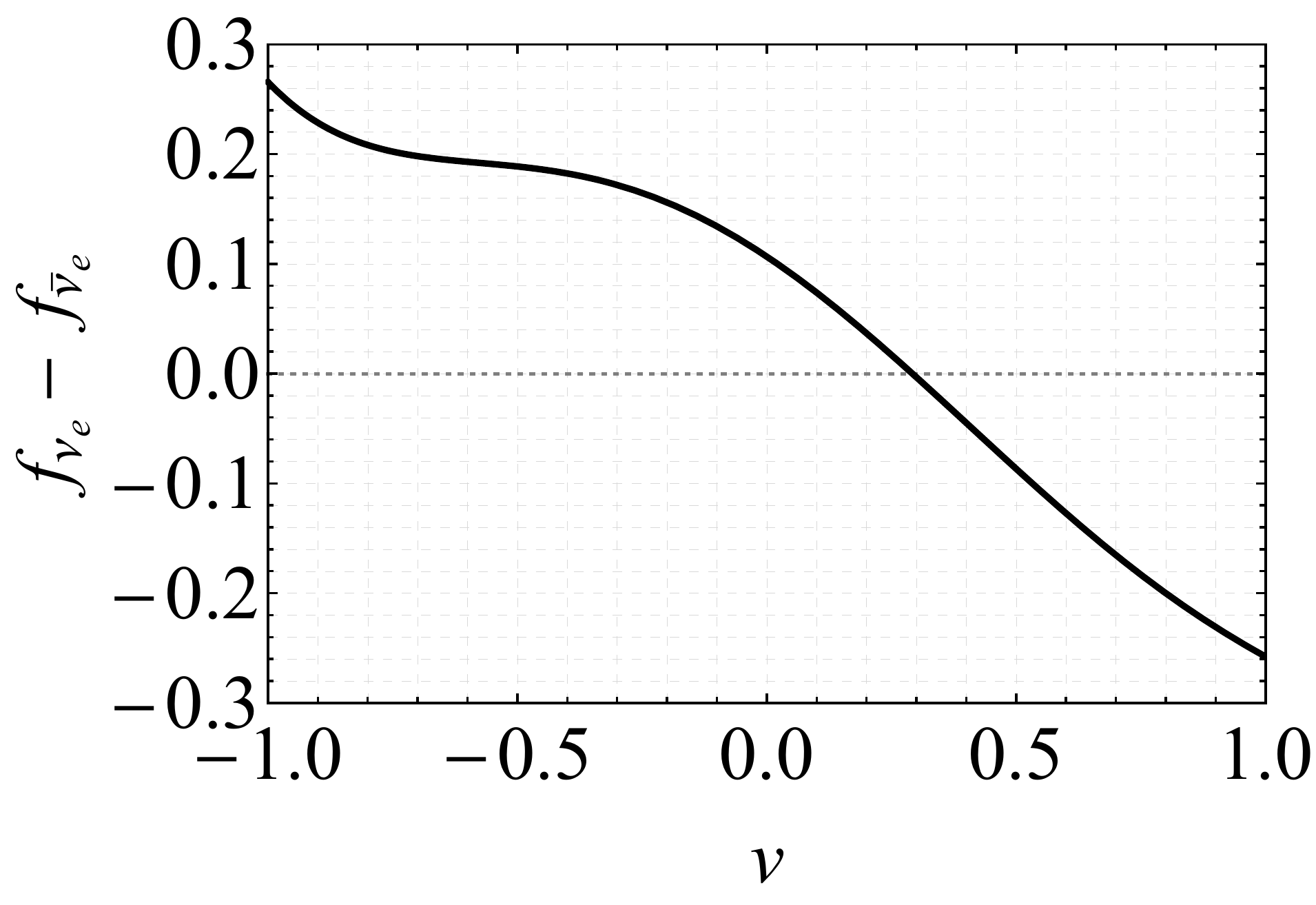}
\end{centering}
\caption{\emph{Left panel:} Zenith-angle distributions of $\nu_e$ and $\bar\nu_e$ inspired by 1D SN models simulated by the Garching group. The relative weights of the fluxes have been changed to generate a smaller asymmetry that leads to a crossing at $v_{\rm c}=0.3$. \emph{Right panel:} Difference of angular spectra of $\nu_e$ and $\bar\nu_e$ showing a crossing at $v_{\rm c}=0.3$.} 
\label{fig:SNspectra}
\vspace{0.5cm}

\begin{centering}
\hspace{0.2in}\includegraphics[width=0.8\columnwidth,height=0.7\columnwidth]{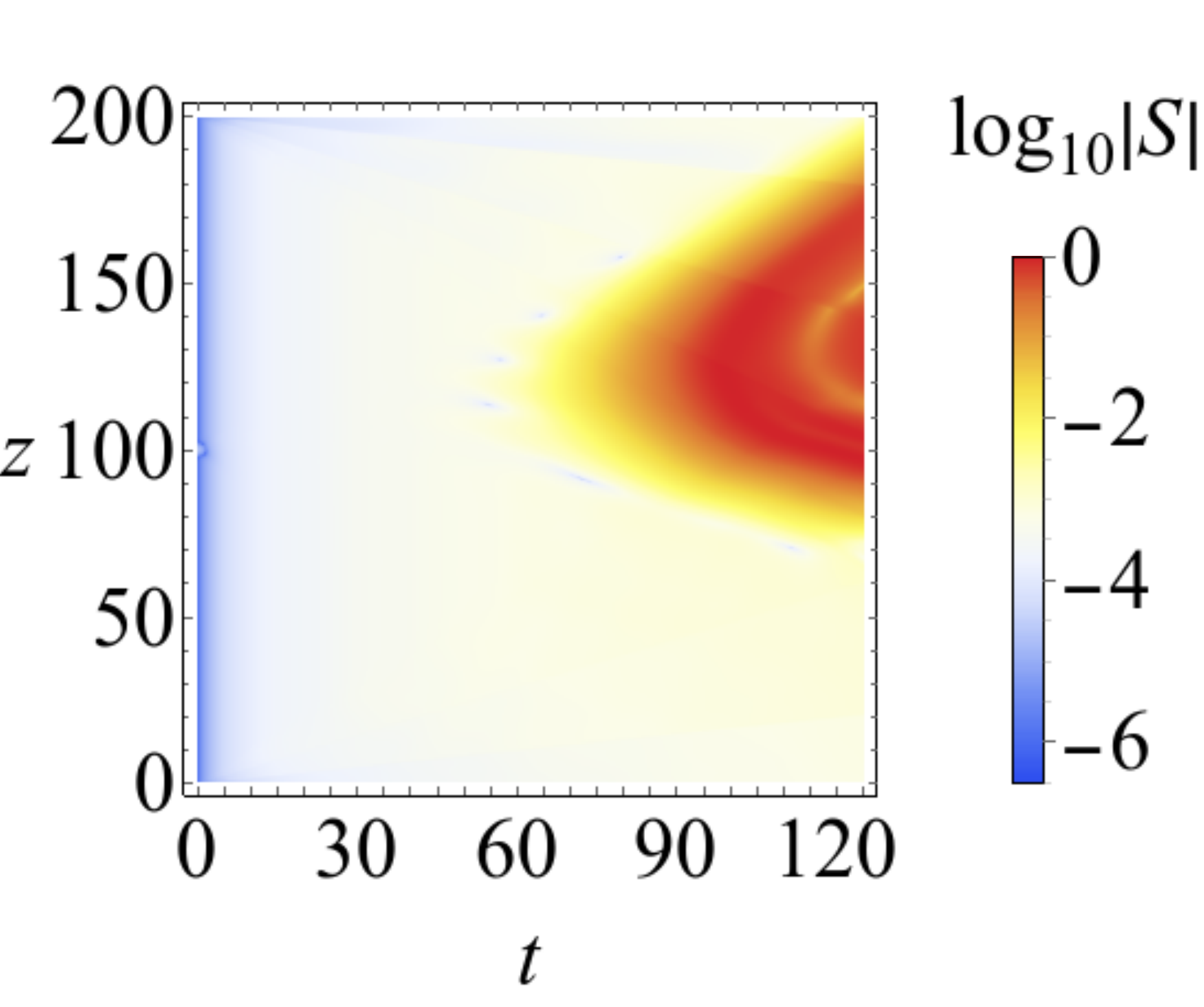}\,
\hspace{1.0cm}
\includegraphics[width=0.95\columnwidth]{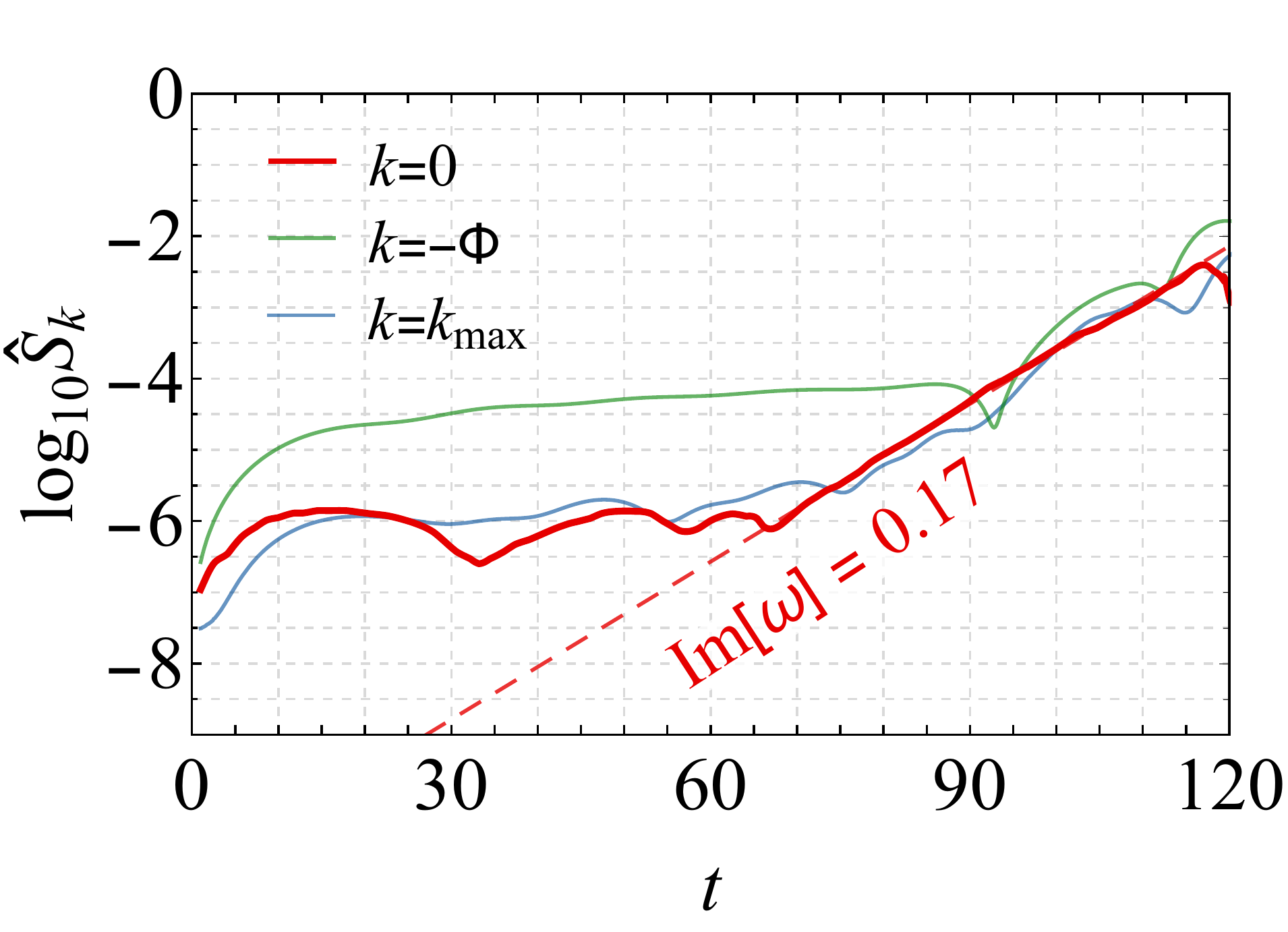}
\end{centering}
\caption{Growth of absolute flavor instability for neutrino angular distributions following SN simulations, as shown in Fig.\,\ref{fig:SNspectra}. \emph{Left panel:} The instability is absolute and spreads over space, without drifting. \emph{Right panel:} The numerically computed growth rate of the zero mode labeled by $k = 0$ (continuous red line), the true homogeneous mode labeled by $k = -\Phi$ (unbroken green line), and the mode with the largest growth rate labeled by $k = k_{\rm max}$ (unbroken blue line). The numerically observed growth rate for the zero mode matches the analytical prediction using the moments (dashed red line).} 
\label{fig:SNinstab}
\end{figure*}  

As a second case we consider a box-like distribution with only forward-going neutrinos $(v>0)$.
In particular, we take 
$(G_{\nu_e}, G_{\bar\nu_e})=(0.6,1.0)$ for $v<0.5$ and $(G_{\nu_e}, G_{\bar\nu_e})=(2.4,1.0)$ for $v>0.5$. In this case
there is a crossing in ELN at $v_{\rm c}=0.5$ but no counter-going neutrinos and we expect a convective instability, where the instability advects away from its point of origin~\cite{Capozzi:2017gqd}.

The numerical solution of the EoMs for this case is shown in left panel of Fig.\,\ref{fig:boxconv}. One finds that the instability is indeed convective, drifting away from its origin at $z \sim 100$ as it grows. In the right panel of Fig.\,\ref{fig:boxconv}, we compare the growth rate for the zero mode (continuous red line), obtained from the numerical solution of the EoMs, with the analytical growth rate ${\rm Im}(\omega)$ predicted by Eq.\,(\ref{eq:growthprox}) (dashed red line).  Clearly, in the exponential growth regime, starting at $t \gtrsim 40$, the agreement is again excellent, with a growth rate ${\rm Im}(\omega) = 0.17$, all the way until saturation. The true homogeneous mode (unbroken green line), and the mode with the maximum growth (unbroken blue line) are also shown for comparison.

As a final example, we consider realistic angular distributions, inspired by 1D SN models simulated by the Garching group, shown in the left panel of Fig.\,\ref{fig:SNspectra}. As discussed in the introduction, most 1D SN simulations do not present a crossing in the ELN, unlike what may be expected in multidimensional SN models. However, the angular distributions are expected to be similar, and we only change the relative weights of $\nu_e$ and $\bar\nu_e$ fluxes within the range predicted by models exhibiting LESA to get a crossing in ELN at $v_{\rm c}=0.3$, as shown in the right panel of Fig.\,\ref{fig:SNspectra}. This ensures that the model shows fast conversions.

In the left panel of Fig.\,\ref{fig:SNinstab} we show the numerical solution of the EoMs for these realistic angular distributions. The dense neutrino cloud has counter-going neutrinos and the instability is absolute, spreading across space at $t \gtrsim 50$, without drifting away completely.  For this model we found, using Eq.\,(\ref{eq:growthprox}), the growth rate to be ${\rm Im}(\omega) = 0.17$. As shown in the right panel, the numerically computed growth rate in the true model for the zero mode (continuous red line) agrees well with the analytical prediction using the moments (dashed red line). The development of the true homogeneous mode (unbroken green line) and the mode with the maximum growth (unbroken blue line) are similar. In fact, in this particular example, the maximum growth rate is roughly the same as that of the zero mode.

\section{Discussion and conclusions}
\label{sec:5}

Fast  flavor conversions can occur close to the neutrino decoupling region in a SN, where ELN angular distributions might have crossings. If these conversions take place, they would bring into question the current paradigm of SN simulations that do not include neutrino flavor conversions in the spectra formation and in SN dynamics~\cite{Dasgupta:2011jf}. Therefore, it is imperative to scan over a large sample of multidimensional SN simulations and search for fast neutrino flavor instabilities. 
Unfortunately, the relevant length scales for fast conversions are much
smaller than the resolution of SN simulations and most multidimensional supernova simulations do not provide detailed neutrino angular distributions, but rather only their integrated moments. Also, numerically identifying the singularities of the Green's function, in order to find all possible instabilities, is difficult and time consuming. 
Therefore, it is perhaps necessary to adopt a schematic implementation of these effects.

We argue that it is possible to analytically calculate the growth of the zero mode, determined by the background matter and neutrino densities, using only the first three moments of the angular distributions. While the zero mode does not necessarily have the largest growth rate, it allows an easy estimate of possible fast flavor conversions in a supernova. Using simple toy examples of box spectra, as well as realistic angular distributions inspired by SN simulations, we have demonstrated that the numerically computed growth rate for the zero mode exactly matches the predictions from the moments of the angular distributions. For completeness, we have also shown the homogeneous and fastest growing modes in all these examples. In the cases we have checked, the zero mode gives a good indication of the timescale over which fast instabilities lead to large flavor conversions.

In a physical situation, triggering of the unstable modes plays an equally important role in determining the instability of the system~\cite{Airen:2018nvp}. Our method however does not take this information into account, and thus one must consider its limitations. It is possible that although the zero mode is unstable, it is not seeded sufficiently. This may lead to the system being stable even when the zero mode is unstable. Of course, it is also possible for the system to be unstable in cases where the zero mode is stable, if a different mode is unstable and suitably excited. Despite these limitations, going beyond which requires information that is not available in contemporary simulations, the proposed method can predict possible flavor instabilities in multi-D simulations with information that is readily available.

We believe that this simplified approach to fast flavor conversions may allow rapid progress in this line of research. Indeed, with our simple recipe given in Sec.\,\ref{sec:3}, it should be possible to perform a preliminary scan for possible fast flavor instabilities in 2D and 3D supernova (and neutron star merger) models. SN simulators are likely to find the computational cost of this method to be significantly lower and might want to use it as a consistency check on their simulations. If unstable cases are found, this would have a profound impact on SN simulations and one would be forced to include the effect of fast conversions into state-of-the-art SN simulations in order to obtain a correct description of the SN dynamics and of the observable SN neutrino fluxes.  

\section*{Acknowledgments}
We thank Francesco Capozzi, Hans-Thomas Janka, Eligio Lisi, Evan O'Connor, and Georg Raffelt for useful discussions and comments on the manuscript. A.M. acknowledges the kind hospitality at TIFR, where part of this work was done.
The work of B.D. is partially supported by the Dept. of Science and Technology of the Govt.\,of India through a Ramanujam Fellowship and by the Max-Planck-Gesellschaft through a Max-Planck-Partnergroup. The work of 
A.M. is supported by the Italian Istituto Nazionale di Fisica Nucleare (INFN) through the ``Theoretical Astroparticle Physics'' project and by Ministero dell'Istruzione, Universit\`a e Ricerca (MIUR).

\bibliographystyle{JHEP}
\bibliography{moments}

\end{document}